\documentclass[12pt,prd,tightenlines,nofootinbib,showpacs,showkeys]{revtex4}
\newcommand{\be}{\begin{equation}}
\newcommand{\ee}{\end{equation}}
\newcommand{\bdis}{\begin{displaymath}}
\newcommand{\edis}{\end{displaymath}}
\newcommand{\bga}{\begin{equation}\begin{gathered}}
\newcommand{\ega}{\end{gathered}\end{equation}}
\usepackage{bm}
\usepackage{graphics}
\usepackage{rotating}
\usepackage{epsfig}
\usepackage{amssymb}
\usepackage{amsmath}
\usepackage{amsthm}
\usepackage{amsfonts}
\usepackage{amssymb}
\usepackage{braket}
\usepackage{graphics}
\usepackage{rotating}
\usepackage{epsfig}
\usepackage{indentfirst}
\usepackage{epstopdf}

\begin{document}
\title{\begin{flushright}{\rm\normalsize SSU-HEP-16/07\\[5mm]}\end{flushright}
Vacuum polarization and quadrupole corrections to the hyperfine splitting of P-states in muonic deuterium}
\author{\firstname{A.~P.} \surname{Martynenko}}
\affiliation{Samara University, Moskovskoye Shosse 34, 443086, Samara, Russia }
\author{\firstname{V.~V.} \surname{Sorokin}}
\affiliation{Samara University, Moskovskoye Shosse 34, 443086, Samara, Russia }

\begin{abstract}
On the basis of quasipotential approach in quantum electrodynamics we calculate vacuum polarization
and quadrupole corrections in first and second orders of perturbation theory in hyperfine structure
of P-states in muonic deuterium. All corrections are presented in integral form and evaluated
analytically and numerically. The obtained results can be used for the improvement of the transition frequencies
between levels 2P and 2S.
\end{abstract}



\maketitle

In last years a significant theoretical interest in the investigation of fine and hyperfine
energy structure of simple atoms is related with light muonic atoms: muonic hydrogen,
muonic deuterium and ions of muonic helium. This is conditioned due to essential progress
achieved by experimental collaboration CREMA (Charge Radius Experiment with Muonic
Atoms) in studies of such simple atoms \cite{crema0,crema1}. In these experiments, the aim is
to clarify the values of the charge radii of nuclei.
The measurement of the transition frequency
($2S^{f=1}_{1/2} - 2P^{f=2}_{3/2}$) leads to a new more precise value of the proton charge radius. For the
first time the hyperfine splitting (HFS) of 2S state in muonic hydrogen was measured in \cite{crema1}. Analogous
measurements in muonic deuterium and muonic helium ions are also carried out \cite{crema2}.
The experiments with muonic hydrogen and deuterium have shown that there are significant discrepancies
between the values of charge radii of the proton and deuteron obtained from experiments
with electronic and muonic atoms.
Successful realization of experimental program is based on precise theoretical
calculations of different corrections to the energy intervals of fine and hyperfine structure
of muonic atoms \cite{borie,borie1,eides,mart1,mart11,miller,pohl1,structure,sgk,udj,carlson,pi,pineda}.
The magnitude of various theoretical contributions to the
energy levels is determined by the values of fundamental physical constants.
The emerging discrepancy requires new analysis of various corrections in the spectrum,
in spite of the fact that the initial estimate of some of them has a small value.
Contributions connected with the structure of nuclei and vacuum polarization
as well as combined corrections have in the calculations the increasing importance. In this work
we investigate a special class of corrections on deuteron structure and vacuum polarization for
hyperfine splitting of P-states in muonic deuterium. In contrast to our previous work \cite{fmms2015},
we use the coordinate representation to calculate the various matrix elements.

Let us consider the HFS of P-states in muonic deuterium. Our approach is based on
quasipotential method in quantum electrodynamics (QED) \cite{fm1999,mart10,mart110},
in which two-particle bound state is described by the Schr\"odinger
equation. Main contribution to hyperfine splitting in muonic deuterium is given by hyperfine part of
the Breit Hamiltoian \cite{fmms2015,mart2,brodsky}:
\begin{equation}
\label{eq:breit}
\Delta V^{hfs}_B(r)=\frac{Z\alpha (1+\kappa_{d})}{2m_1 m_2 r^3}\bigl[ 1+\frac{m_1 \kappa_{d}}
{m_2 (1+\kappa_{d})}\bigr](\boldsymbol L\cdot\boldsymbol s_2)-\frac{Z\alpha (1+\kappa_{d})
(1+a_{\mu})}{2m_1 m_2 r^3}\bigl[(\boldsymbol s_1\cdot \boldsymbol s_2)-
3(\boldsymbol s_1\cdot\boldsymbol n)(\boldsymbol s_2\cdot \boldsymbol n)\bigr],
\end{equation}
where $m_1$, $m_2$ are muon and deuteron masses respectively, $\kappa_d$, $a_{\mu}$ are anomalous
magnetic moments of deuteron and muon, $\boldsymbol L$, $\boldsymbol s_1$ are orbital momentum
and spin of muon, $\boldsymbol s_2={\bf I}$ is the deuteron spin, $\boldsymbol n = \boldsymbol r /r.$
This operator doesn't commute with the operator of total angular momentum of muon
$\boldsymbol J=\boldsymbol L+\boldsymbol s_1$, which leads to non-zero off-diagonal matrix elements.
The Coulomb wave function for 2P-state has the following form:
\begin{equation}
\label{eq:psi}
\Psi_{2P}(\boldsymbol r)=\frac{1}{2\sqrt{6}}W^{\frac{5}{2}}re^{-\frac{Wr}{2}}Y_{1m}(\theta,\phi),~~~W=\mu Z \alpha.
\end{equation}
Averaging \eqref{eq:breit} over wave function \eqref{eq:psi} we obtain the contributions of order $\alpha^4$ to HFS of P-states
which are written analytically in \cite{fmms2015} and numerically in Table~\ref{tbl1}. The one-loop vacuum polarization correction
to the operator \eqref{eq:breit} has the following form \cite{mart3}:
\begin{equation}
\label{eq:1gammavppot}
\Delta V^{hfs}_{1\gamma,vp}(r)=\frac{Z\alpha^2(1+\kappa_d)}{6\pi m_1 m_2 r^3}
\int_1^\infty\rho(\xi)d\xi e^{-2m_e\xi r} \biggl\{\biggl( 1+\frac{m_1\kappa_d}
{m_2 (1+\kappa_d)} \biggl)(\boldsymbol L\cdot\boldsymbol s_2)
(1+2m_e\xi r)-
\end{equation}
\begin{displaymath}
(1+a_\mu)\biggl(4m_e^2 \xi^2 r^2[(\boldsymbol s_1\cdot \boldsymbol s_2)-
(\boldsymbol s_1\cdot\boldsymbol n)(\boldsymbol s_2\cdot \boldsymbol n)]+
(1+2 m_e \xi r)[(\boldsymbol s_1\cdot \boldsymbol s_2)-
3(\boldsymbol s_1\cdot\boldsymbol n)(\boldsymbol s_2\cdot \boldsymbol n)]\biggl)\biggl\}.
\end{displaymath}
The contribution of \eqref{eq:1gammavppot} to hyperfine splitting is given by integral expression:
\begin{equation}
\label{eq:1gammavp}
E^{hfs}_{1\gamma,vp}(r)=\frac{\alpha^4 \mu^3(1+\kappa_d)}{24m_1 m_2 r^3} \frac{\alpha}{6\pi}\int_1^\infty\rho(\xi)d\xi \int_0^\infty x dx e^{-x[1+\frac{2m_e\xi}{W}]} \biggl[\biggl( 1+\frac{m_1\kappa_d}{m_2 (1+\kappa_d)} \biggl)\times
\end{equation}
\begin{displaymath}
\overline{T_1}
(1+\frac{2m_e\xi}{W} x)-(1+a_\mu)\biggl(\frac{4m_e^2 \xi^2 x^2}{W^2}\overline{T_3}+(1+\frac{2 m_e \xi}{W} x)\overline{T_2}\biggl)\biggl],
\end{displaymath}
where we introduce the following designations for the operators $T_i$ (i=1,2,3):
\begin{equation}
\label{eq:angldes}
T_1=(\boldsymbol L\cdot\boldsymbol s_2),~T_2=\biggl[(\boldsymbol s_1\cdot \boldsymbol s_2)-
3(\boldsymbol s_1\cdot\boldsymbol n)(\boldsymbol s_2\cdot \boldsymbol n)\biggl],~
T_3=\biggl[(\boldsymbol s_1\cdot \boldsymbol s_2)-(\boldsymbol s_1\cdot\boldsymbol n)(\boldsymbol s_2\cdot \boldsymbol n)\biggl].
\end{equation}
To calculate matrix elements $\overline{T_i}$ we use the atomic wave function
\begin{equation}
\label{eq:atomic}
\Psi_{FM}=\sum_{m}C(IjF;M-m,m)\psi_{IM-m}\psi_{jm},
\end{equation}
where $M$ is the projection of total momentum ${\bf F}$ on the z-axis, $m$ is the projection of total
muon momentum ${\bf j}$ on the z-axis.
For the calculation diagonal and off-diagonal matrix elements we can use the Wigner-Eckart theorem which allows
to express the initial matrix element of a scalar product of two rank 1 irreducible operators $T^1$ and
$T^2$ through the reduced matrix
elements \cite{sobelman,quad}:
\begin{equation}
\label{eq:theorem}
<j'IF|(T^1\cdot T^2)|JIF>=(-1)^{I+j'-F}W(jIj'I;F1)<j'||T^1||j><I||T^2||I>,
\end{equation}
where the Racah coefficients are connected with 6J-symbols by the following relation:
\begin{equation}
\label{eq:R6J}
W(j_1j_2j_5j_4;j_3j_4)=(-1)^{-j_1-j_2-j_4-j_5}
\left\{
\begin{array}{ccc}
j_1&j_2&j_3\\
j_4&j_5&j_6\\
\end{array}
\right\}.
\end{equation}
So, for example, for the first matrix element $\overline T_1$ we have:
\begin{equation}
\label{eq:t1}
<j'IF|({\bf L}{\bf s}_2|jIF>=(-1)^{j'-j-F-I+l+3/3}\sqrt{(2j+1)(2j'+1)}
\left\{
\begin{array}{ccc}
j&I&F\\
I&j'&1\\
\end{array}
\right\}\times
\end{equation}
\begin{displaymath}
\left\{
\begin{array}{ccc}
l&j'&\frac{1}{2}\\
j&l&1\\
\end{array}
\right\}
<l||L||l><I||I||I>=(-1)^{j'-j-F-I+l+3/3}\sqrt{(2j+1)(2j'+1)}\times
\end{displaymath}
\begin{displaymath}
\sqrt{I(I+1)(2I+1)l(l+1)(2l+1)}
\left\{
\begin{array}{ccc}
j&I&F\\
I&j'&1\\
\end{array}
\right\}
\left\{
\begin{array}{ccc}
l&j'&\frac{1}{2}\\
j&l&1\\
\end{array}
\right\}.
\end{displaymath}

Calculating coefficients in right part of \eqref{eq:t1} we obtain numerical value $\overline T_1$ in the case
of P-states. Similar evaluation of other matrix elements $\overline T_i$ gives the following result for
diagonal matrix elements with $j=j'=1/2$ and $j=j'=3/2$ and off-diagonal matrix elements with $j=1/2$, $j'=3/2$:
\begin{equation}
\label{eq:tton}
\overline{T_1}\bigl(\frac{1}{2},\frac{1}{2}\bigr)=-\overline{T_2}=-2\overline{T_3}=-\frac{4}{3}
\delta_{F\frac{1}{2}}+\frac{2}{3}\delta_{F\frac{3}{2}},~
\overline{T_1}\bigl(\frac{3}{2},\frac{3}{2}\bigr)=5\overline{T_2}=\frac{5}{2}\overline{T_3}=-\frac{5}{3}
\delta_{F\frac{1}{2}}-
\frac{2}{3}\delta_{F\frac{3}{2}}+\delta_{F\frac{5}{2}},
\end{equation}
\begin{equation}
\label{eq:ttoff}
\overline T_1\bigl(\frac{1}{2},\frac{3}{2}\bigr)=2\overline T_2\bigl(\frac{1}{2},\frac{3}{2}\bigr)=-2
\overline T_3\bigl(\frac{1}{2},\frac{3}{2}\bigr)=
\Biggl\{
\begin{array}{cc}
-\frac{\sqrt{2}}{3}, & F=\frac{1}{2}\\
-\frac{\sqrt{5}}{3}, & F=\frac{3}{2}\\
\end{array}
.
\end{equation}
The integration in \eqref{eq:1gammavp} is performed analytically over $x$ and numerically over $\xi$.
Numerical results are presented in Table \ref{tbl1}.
\begin{table}[htbp]
\caption{Numerical values of corrections to $2P$-state hyperfine structure}
\label{tbl1}
\begin{center}
\centering
\begin{tabular}{|c|c|c|c|c|c|c|c|}   \hline
Contribution   & $2^2P_{1/2}$& $2^4P_{1/2}$& $2^2P_{3/2}$& $2^4P_{3/2}$& $2^6P_{3/2}$&$2^2P_{1/2 \rightarrow 3/2}$&$2^4P_{1/2 \rightarrow 3/2}$\\
             & $(\mu eV)$ &$(\mu eV)$ &$(\mu eV)$ &$(\mu eV)$ & $(\mu eV)$&$(\mu eV)$&$(\mu eV)$ \\   \hline
Leading order& -1380.3359             & 690.1679           & 8162.2889           & 8583.2316 & 9284.8027      & -126.0372 & -199.2824\\
$\alpha^4$ contribution   &&&&&&&\\ \hline
Relativistic                        & -0.1676  & 0.0838  & -0.0125 & -0.0050  & 0.0075  & -0.0043  & -0.0067\\
correction of order $\alpha^6$  &&&&&&&\\ \hline
VP corrections      & -1.0706                & 0.5353             & -0.2802  & -0.1121 & 0.1681         & -0.1437  & -0.2271\\
of order $\alpha^5$ &   &   &    &   &   &   &   \\  \hline
VP corrections      & -0.0011                & 0.0005             & -0.0014             & -0.0006  & 0.0008   & 0.00005   & 0.0001\\
of order $\alpha^6$  &   &   &   &   &   &   &   \\   \hline
Quadrupole    & 0  & 0  & 434.2329 & -347.3863  & 86.8466  & 614.0980   & -194.1948\\
correction of order $\alpha^4$  &&&&&&&\\  \hline
Quadrupole and VP                   & 0                      & 0  & 0.2438  & -0.1950   & 0.0488    & 0.3447  & -0.1090\\
correction of order $\alpha^5$  &&&&&&&\\
in $1\gamma$ interaction &  &  &  &  &  &  &  \\  \hline
Quadrupole and VP              & 0                      & 0 & 0.1122   & -0.0898   & 0.0224  & 0.1587   & -0.0502\\
correction of order $\alpha^5$  &&&&&&&        \\
in second order PT  &  &  &  &  &  &  &  \\  \hline
Total value     &  -1381.5752  & 690.7876   & 8596.5838   & 8235.4428  & 9371.8969 & 488.4164  & -393.8702\\ \hline
\end{tabular}
\end{center}
\end{table}

For two-loop vacuum polarization contributions into a potential (loop after loop term and 2-loop term)
we have the following expressions which contain the same tensor operators
$T_i$ as above \cite{mart2,mart3}:
\begin{equation}
\label{eq:vpvp}
\Delta V^{hfs}_{1\gamma,vp-vp}(r)=\frac{Z\alpha(1+\kappa_d)}{2m_1 m_2 r^3}\bigl( \frac{\alpha}{3\pi}\bigr)^2
\int_1^\infty\rho(\xi)d\xi\int_1^\infty\rho(\eta)d\eta \frac{1}{\xi^2-\eta^2}\bigl[\bigl(1+\frac{m_1\kappa_d}{m_2 (1+\kappa_d)}\bigr)(\boldsymbol L\cdot\boldsymbol s_2)\times
\end{equation}
\begin{displaymath}
[\xi^2(1+2m_e \xi r)
e^{-2m_e \xi r}-\eta^2(1+2m_e \eta r)e^{-2m_e \eta r}]-(1+a_\mu)\biggl(4m_e^2 r^2[\xi^4 e^{-2 m_e \xi r}-\eta^4 e^{-2 m_e \eta r}]\times
\end{displaymath}
\begin{displaymath}
[(\boldsymbol s_1\cdot \boldsymbol s_2)-
(\boldsymbol s_1\cdot\boldsymbol n)(\boldsymbol s_2\cdot \boldsymbol n)]+[\xi^2(1+2m_e \xi r)e^{-2m_e \xi r}-\eta^2(1+2m_e \eta r)e^{-2m_e \eta r}]
[(\boldsymbol s_1\cdot \boldsymbol s_2)-3(\boldsymbol s_1\cdot\boldsymbol n)(\boldsymbol s_2\cdot \boldsymbol n)]\bigr)\bigr],
\end{displaymath}
\begin{equation}
\label{eq:2loopvp}
\Delta V^{hfs}_{2-loop~vp}(r)=\frac{Z\alpha(1+\kappa_d)}{2m_1 m_2 r^3}
\frac{2}{3}\bigl(\frac{\alpha}{\pi}\bigr)^2\int_0^1 \frac{f(v)dv}{1-v^2} e^{-\frac{2m_e r}
{\sqrt{1-v^2}}}\bigl[\bigl( 1+\frac{m_1\kappa_d}{m_2 (1+\kappa_d)} \bigr)
\bigl[1+\frac{2m_e r}{\sqrt{1-v^2}} \bigr](\boldsymbol L\cdot\boldsymbol s_2)-
\end{equation}
\begin{displaymath}
(1+a_\mu)\biggl(\frac{4m_e^2 r^2}{1-v^2}[(\boldsymbol s_1\cdot \boldsymbol s_2)-
(\boldsymbol s_1\cdot\boldsymbol n)(\boldsymbol s_2\cdot \boldsymbol n)]+
\bigl(1+\frac{2m_e r}{\sqrt{1-v^2}} \bigr)[(\boldsymbol s_1\cdot \boldsymbol s_2)-
3(\boldsymbol s_1\cdot\boldsymbol n)(\boldsymbol s_2\cdot \boldsymbol n)]\bigr)\bigr].
\end{displaymath}

After averaging \eqref{eq:vpvp} and \eqref{eq:2loopvp} over wave functions we obtain numerical
values of corresponding corrections to the HFS
that are included in Table~\ref{tbl1}. Numerically they are extremely small.
Muonic VP correction of order $\alpha^6$ can de derived by means of simple replacement $m_e$
to $m_1$ in \eqref{eq:1gammavp}.
One-loop VP contribution to the HFS in second order perturbation theory (SOPT) has the
following general form \cite{mart2,mart3}:
\begin{equation}
\label{eq:soptmainf}
\Delta E^{hfs}_{SOPT~vp}=2<\psi|\Delta V^C_{vp}\cdot \tilde G\cdot\Delta
V_B^{hfs}|\psi>,
\end{equation}
where $\Delta V^C_{vp}(r)$ is the Coulomb potential modified by the vacuum polarization.
The Coulomb Green's function $\tilde G$ for $2P$-state was obtained in \cite{hameka}.
Substituting \eqref{eq:breit} and $\Delta V^C_{vp}(r)$ into \eqref{eq:soptmainf} we get integral
expression for the VP correction in SOPT:
\begin{equation}
\label{eq:sopt}
\Delta E_{SOPT,vp}^{hfs}=\frac{\alpha^5\mu^3(1+\kappa_d)}{1296\pi m_1 m_2}
\int_1^\infty\rho(\xi)d\xi\int_0^\infty dx e^{-x(1+\frac{2m_e\xi}{W})} \int_0^\infty\frac{e^{-x'}dx'}{x'^2}g(x,x')\times
\end{equation}
\begin{displaymath}
\biggl[\biggl(1+\frac{m_1\kappa_{d}}{m_2(1+\kappa_{d})}\biggr)\overline{T_1}-(1+a_{\mu})\overline{T_2} \biggr].
\end{displaymath}
The integration in \eqref{eq:sopt} is performed analytically over $x$, $x'$ and numerically over $\xi$.
For two-loop contributions in second order PT we use the potential \eqref{eq:1gammavppot} and the VP modifications
of the Coulomb potential from \cite{mart2,mart3}. Corresponding numerical results are included in Table~\ref{tbl1}.
The contribution of the VP of order $\alpha^6$ to the HFS of P-states in third order PT is estimated using the known
formulas. Omitting here intermediate analytical expressions (see \cite{mart1}) we have included numerical value in Table~\ref{tbl1}.

The deuteron has a non-zero quadrupole moment which leads to additional quadrupole interaction correction of order $\alpha^4$
to hyperfine structure of P-states \cite{fmms2015}. We turn right to the calculation of the quadrupole corrections
for the effects of vacuum polarization. To construct the required interaction operator we use the multipole expansion
of the Coulomb potential, taking into account the effect of vacuum polarization:
\begin{equation}
\label{eq:multipole}
V^C_{vp}(r)=-Ze^2\frac{\alpha}{3\pi}\int_1^\infty\rho(\xi)d\xi\int\frac{\rho(r')d{\bf r}'}{|{\bf r}-{\bf r}'|}e^{-2m_e\xi|{\bf r}-{\bf r}'|}=
\end{equation}
\begin{displaymath}
-Ze^2\frac{\alpha}{3\pi r}\int_1^\infty\rho(\xi)e^{-2m_e\xi r}d\xi\int\rho(r')d{\bf r}'\bigl[P_0f_0(r,r',\xi)+
P_1f_1(r,r',\xi)+P_2f_2(r,r',\xi)+...\bigr],
\end{displaymath}
where $P_n(\cos\theta)$ are the Legendre polynomials, $\rho(r')$ is the nuclear density,
\begin{equation}
\label{eq:leg}
f_0(r,r',\xi)=1+\frac{r'^2}{r^2}\frac{4m_e^2\xi^2r^2}{6},
f_1(r,r',\xi)=\frac{r'}{r}(1+2m_e\xi r), f_2(r,r',\xi)=\frac{r'^2}{r^2}(1+2m_e\xi r+\frac{4m_e^2\xi^2r^2}{3}).
\end{equation}
A separation of the muon and nuclear variables in \eqref{eq:multipole} can be made using the addition theorem
for spherical harmonics. As a result, the expression \eqref{eq:multipole} is converted into a sum of scalar
products of irreducible tensors.
The third term in square brackets in \eqref{eq:multipole} is determined by the nuclear quadrupole momentum.
Then the energy of quadrupole interaction with vacuum polarization correction can be presented in the form:
\begin{equation}
\label{eq:qvp}
\Delta E_{vp}^Q=-Ze^2<FMIj'|Q_{20}(d)\cdot T_{20}^{vp}(\mu)|FMIj>=
\end{equation}
\begin{displaymath}
-Ze^2(-1)^{I+j'-F}W(j I j' I;F 2)<j'||T_{20}^{vp}(\mu)||j><I||Q_{20}(d)||I>,
\end{displaymath}
where nuclear quadrupole tensor and a tensor of muon cloud of rank 2 are equal correspondingly
\begin{equation}
\label{eq:q20}
Q_{20}(d)=\sqrt{\frac{4\pi}{5}}r'^2Y_{20}(\theta',\phi')=\frac{3z'^2-r'^2}{2},
\end{equation}
\begin{equation}
\label{eq:t20}
T_{20}^{vp}(\mu)=\frac{\alpha}{3\pi}\int_1^\infty\rho(\xi)d\xi \frac{e^{-2m_e\xi r}}{r^3}
\left(1+2m_e\xi r+\frac{4m_e^2\xi^2 r^2}{3}\right)\sqrt{\frac{4\pi}{5}}Y_{20}(\theta,\phi).
\end{equation}
The nuclear quadrupole momentum $Q$ is determined by reduced matrix element of deuteron tensor
$Q_{20}(d)$ \cite{sobelman}:
\begin{equation}
\label{eq:quadrel1}
<I||Q_{20}(d)||I>=\frac{Q}{2} \biggl[ \biggl(
\begin{array}{ccc}
I & 2 & I \\
-I & 0 & I
\end{array}
\biggr) \biggr]^{-1},
\end{equation}
In turn, the reduced matrix element of muon tensor differs from the case of the quadrupole
correction of the leading order only by the value of radial integral:
\begin{equation}
\label{eq:quadrel2}
<j'||T_{20}^{vp}(\mu)||j>=\sqrt{2j+1}\sqrt{2j'+1}(-1)^{j'+1/2}\biggl(
\begin{array}{ccc}
j' & 2 & j \\
\frac{1}{2} & 0 & -\frac{1}{2}
\end{array}
\biggl)<\frac{e^{-2m_e\xi r}}{r^3}\bigl(1+2m_e\xi r+\frac{4m_e^2\xi^2 r^2}{3}\bigr)>,
\end{equation}

Substituting \eqref{eq:quadrel1} and \eqref{eq:quadrel2} into \eqref{eq:qvp} we obtain
the quadrupole interaction contribution with the account of vacuum polarization effects
in calculating both the diagonal and off-diagonal matrix elements in the form:
\begin{equation}
\label{eq:quadform}
\Delta E_{vp}^Q=(-1)^{j'+1/2-F-j}\biggl\{
\begin{array}{ccc}
j & I & F \\
I & j' & 2
\end{array}
\biggl\}\frac{Q}{2} \biggl[ \biggl(
\begin{array}{ccc}
I & 2 & I \\
-I & 0 & I
\end{array}
\biggl)\biggl]^{-1}\times
\end{equation}
\begin{displaymath}
\notag\times\sqrt{2j+1}\sqrt{2j'+1}\biggl(
\begin{array}{ccc}
j' & 2 & j \\
\frac{1}{2} & 0 & -\frac{1}{2}\\
\end{array}
\biggr)
<\frac{Z\alpha^2 e^{-2m_e\xi r}}{r^3}\bigl(1+2m_e\xi r+\frac{4m_e^2\xi^2 r^2}{3}\bigr)>.
\end{displaymath}

For diagonal and off-diagonal matrix elements we get following analytical integral expressions
with subsequent numerical evaluation:
\begin{equation}
\label{eq:qqvp}
\Delta E^Q_{vp}(j'=3/2,j=3/2)=\frac{\mu^3\alpha(Z\alpha)^4Q}{36\pi}\int_1^\infty\frac{(5a^2+8a+4)}{(a+2)^4}\rho(\xi)d\xi
\left[\delta_{F\frac{1}{2}}-\frac{4}{5}\delta_{F\frac{3}{2}}+\frac{1}{5}\delta_{F\frac{5}{2}}\right]=
\end{equation}
\begin{equation*}
=\left[\delta_{F\frac{1}{2}}-\frac{4}{5}\delta_{F\frac{3}{2}}+\frac{1}{5}\delta_{F\frac{5}{2}}\right]\times 0.2441~\mu eV,
\end{equation*}
\begin{equation}
\label{eq:qqvp1}
\Delta E^Q_{vp}(j'=1/2,j=3/2)=\frac{\mu^3\alpha(Z\alpha)^4Q}{36\pi}\int_1^\infty\frac{(5a^2+8a+4)}{(a+2)^4}\rho(\xi)d\xi
\left[\sqrt{2}\delta_{F\frac{1}{2}}-\frac{1}{\sqrt{5}}\delta_{F\frac{3}{2}}\right]=
\end{equation}
\begin{equation*}
=\left[\sqrt{2}\delta_{F\frac{1}{2}}-\frac{1}{\sqrt{5}}\delta_{F\frac{3}{2}}\right]\times 0.2441~\mu eV,
\end{equation*}
where the value of deuteron quadrupole moment is equal to $Q=0.285783(30)~fm^2$ \cite{quadval},
$a=4m_e\xi/\mu\alpha$. Note that numerical
coefficients in \eqref{eq:qqvp} and \eqref{eq:qqvp1} coincide.
The quadrupole and VP corrections to diagonal and off-diagonal matrix elements in second order PT
which can be calculated as in \eqref{eq:soptmainf} are equal to
\begin{equation}
\label{eq:quadVPSOPTdiag}
E^{Q}_{sopt,vp}(j'=3/2,j=3/2)=\left[\delta_{F\frac{1}{2}}-\frac{4}{5}\delta_{F\frac{3}{2}}+\frac{1}{5}\delta_{F\frac{5}{2}}\right]\times  0.1122~\mu eV,
\end{equation}
\begin{equation}
\label{eq:quadVPSOPToffdiag}
E^{Q}_{sopt, vp}(j'=1/2,j=3/2)=\left[\sqrt{2}\delta_{F\frac{1}{2}}-\frac{1}{\sqrt{5}}\delta_{F\frac{3}{2}}\right]\times  0.1122~\mu eV.
\end{equation}

In this work, we have continued to study the P-states HFS in muonic deuterium, which began in \cite{fmms2015}.
If in \cite{fmms2015} a method for HFS calculating  of P-levels in momentum representation was formulated, in this paper
we have used the formalism of irreducible tensor operators in the calculation of corrections in the coordinate representation.
Both representations are complementary to each other, so that the calculation of the matrix elements of the operators 
in both representations is in our opinion useful.
The results obtained here and in \cite{fmms2015} are in agreement, but
we are able to correct technical inaccuracy in the calculation of the off-diagonal matrix elements in \cite{fmms2015}
(see \eqref{eq:qqvp1}).
We present corrections in integral form and evaluate them numerically. After diagonalization of the results from the energy matrix in Table~\ref{tbl1}
we obtain final values of $2P$-state hyperfine structure in muonic deuterium: $E_{1/2}^{F=1/2}=-1405.4254~\mu eV$, $E_{1/2}^{F=3/2}=670.2812~\mu eV$, $E_{3/2}^{F=1/2}=8620.4340~\mu eV$, $E_{3/2}^{F=3/2}=8255.9492~\mu eV$, $E_{3/2}^{F=5/2}=9371.8969~\mu eV$.
The detailed calculation of 2P-state HFS in muonic deuterium was performed in~\cite{borie}, where first order PT vacuum polarization corrections were included. Vacuum polarization corrections in \cite{borie} were evaluated approximately thus they differ from our values \eqref{eq:1gammavp} by $\sim 30\%$. Other differences are connected with second order PT $\alpha^5$ and $\alpha^6$ corrections. Obtained results can be used for improved estimates of transition frequencies between $2P$ and $2S$ states regarding to the CREMA experiments.

We are grateful to F.~Kottmann, R.~Pohl and B.~Franke for useful discussions and information about recent experimental
results of the CREMA collaboration.
The work is supported by the Russian Foundation for Basic Research (grant 16-02-00554) and
the Ministry of Education and Science of Russia under grant No.~1394.

\bibliographystyle{aipproc}

\end{document}